\documentstyle[11pt,epsfig]{article}
\begin{document}
\title{\bf Gravitomagnetism and Non--commutative Geometry}
\author{Behrooz Malekolkalami${}^1$\thanks{E-mail: B.malakolkalam@uok.ac.ir} \ and
        Mehrdad Farhoudi${}^2$\thanks{E-mail: m-farhoudi@sbu.ac.ir}\\
        {\small ${}^1$Department of Physics,
        University of Kurdistan, P.O. Box 66177-15175, Sanandaj, Iran}\\
        {\small ${}^2$Department of Physics,
        Shahid Beheshti University G.C., Evin, Tehran 19839, Iran} }
\date{\small November 4, 2013}
\maketitle
\begin{abstract}
\noindent
 Similarity between the gravitoelectromagnetism and the
electromagnetism is discussed. We show that the gravitomagnetic
field (similar to the magnetic field) can be equivalent to the
non--commutative effect of the momentum sector of the phase space
when one maintains only the first order of the non--commutative
parameters. This is performed through two approaches. In one
approach, by employing the Feynman proof, the existence of a
Lorentz--like force in the gravitoelectromagnetism is indicated.
The appearance of such a force is subjected to the slow motion and
the weak field approximations for stationary fields. The analogy
between this Lorentz--like force and the motion equation of a test
particle in a non--commutative space leads to the mentioned
equivalency. In fact, this equivalency is achieved by the
comparison of the two motion equations. In the other and quietly
independent approach, we demonstrate that a gravitomagnetic
background can be treated as a Dirac constraint. That is, the
gravitoelectromagnetic field can be regarded as a constrained
system from the sense of the Dirac theory. Indeed, the application
of the Dirac formalism for the gravitoelectromagnetic field
reveals that the phase space coordinates have non--commutative
structure from the view of the Dirac bracket. Particularly, the
gravitomagnetic field as a weak field induces the non--trivial
Dirac bracket of the momentum sector which displays the
non--commutativity.
\end{abstract}
\bigskip
\section{Introduction}
\indent

The analogy between gravitation and electromagnetism originates in
the similarity between the Coulomb law of electricity and the
Newton law of gravitation. Actually, the analogous idea of the
electric theory and the Newtonian gravitational theory, which
inspires a Maxwell--type of gravitational theory, is dated back to
the second half of the nineteenth century~\cite{1}--\cite{4}. The
motivation that the motion of a mass can generate a field
analogous to the magnetic field is emerged from the fact that the
magnetic field is produced by the electric current. Indeed, the
mass current produces a field called gravitomagnetic ({\bf GM})
field. The introduction of  GM field as an analog to the magnetic
field comes from the need for knowing the force exerted by a
moving body based on the intriguing interplay between geometry and
dynamics, as emphasized by Sciama~\cite{5}. Of course, such an
analogy is necessarily incomplete, for instance, unlike the
electric charge, the mass charges are~not invariant nor
additive~\cite{51}. Thus, a more natural framework where this
topic can be developed, is the theory of general relativity. The
first general investigation of the GM field within this theory is
due to Thirring~\cite{52}. In fact, the introduction of  GM field
is unavoidable when one brings the Newtonian gravitational theory
and the Lorentz invariance together in a consistent framework.
This effect is the usual Maxwellian feature which has Machian
root, see, e.g., Ref.~\cite{6}. Dynamical equations for a weak
gravitational field, similar to the Maxwell equations, has been
deduced by the parameterized post Newtonian formalism~\cite{7}.
Also, more attention has been made in the analogy between general
relativity and electromagnetism for slowly motion in weak
gravitational fields~\cite{8}.

There are other dissimilarities in this analogy. For example, a
negative mass charge has~not been detected yet\footnote{However,
there are some interpretation on the cosmological constant
(associated with dark energy) that acts effectively as a repulsive
(gravity) force, particularly on the large scale. In this issue,
see, e.g., Refs.~\cite{Mannheim00,bfgt01,Weinberg05}.}\
 and, the
like mass charges attract rather than repelling each others. Also,
there are issues which partially are stated as the weak version of
the equivalence principle, that is, the gravitational field is
coupled to everything and/or all forms of energy act as sources of
the gravitational field. On the other hand, there is another
(outward) type of similarity between the GM and magnetic fields.
This similarity is related to the effect of non--commutative ({\bf
NC}) parameters on the motion equation of a particle in a NC
space.

The canonical coordinates and momenta, as independent variables,
have equal status and can be designated as the physical
quantities. They are both needed to describe the motion of the
system in the Hamiltonian formulation, and works have been
performed in devising schemes that result in entirely symmetric
equations. That is, the distinction between these two classes of
variables is basically one of nomenclature. In this aspect, the NC
geometry has a genuine base and can be viewed as a manifestation
of this respect, in which it presents a newer and perhaps deeper
insight into the physical contents of the nature (phenomena).
Indeed, when the geometry of the phase space is NC, the motion
equation contains terms including the NC parameters which appear
as additional force terms with respect to the usual space. Also,
when in the usual space, an electromagnetic field is present, the
Lorentzian force gives the motion equation of a moving charged
particle. Under special conditions, by comparing these motion
equations (in the NC and usual spaces), one can deduce that the
effect of the NC parameters on the motion equation is equivalent
to the effect of the magnetic field in the usual space. This
equivalency has been demonstrated in the literature, e.g., in the
classical~\cite{9} and in the quantum~\cite{10} perspectives.

In this work, we purpose to investigate the same equivalency for
the GM field and the NC effects, and we will show that this is
possible if a Lorentz--like force exists for the
gravitoelectromagnetic~({\bf GEM}) field. Actually, such
possibility can be demonstrated by employing the Feynman
proof~({\bf FP}) for the gravitational field. In 1948, Feynman
gave a proof\footnote{This proof was never published by
Feynman~\cite{101}.}\
 of the Maxwell equations by assuming
only the Newton law of motion and the commutation relation between
the position and the velocity of a single non--relativistic
particle. Interestingly, his proof leads to the relativistic
result, that is the existence of a Lorentz--like force for the
motion of a particle. Also, a more important result of the proof
is that, the corresponding electric and magnetic like fields
satisfy the homogenous Maxwell equations. A glance at the proof
illustrates that, it is~not restricted only to the electromagnetic
interactions, but also includes other interactions, among them,
the gravitational one. Of course, the implications of the FP to
the GM interaction is subjected to special conditions~\cite{21}.

Therefore, in this approach by using the FP, we first derive the
motion equation of a particle in the presence of the GM field and
then, compare this equation with the equation motion in the NC
space in order to illustrate the mentioned equivalency. Note that,
in this approach, we deal with two different spaces, where in one
of them (the usual space) an external GM field is present, and in
the other one (the NC space) there is~not. Incidentally, the
equations of motion must be written in the canonical (Hamiltonian)
form, as the functions\rlap,\footnote{The functions that appear in
the definition of the NC product.}\
  in the NC frame, are defined on a relevant phase space. Also,
we consider the non--commutativity in the both sectors of the
phase space, namely the space sector and the momentum sector.

On the other hand, there is an another straightforward
approach~\cite{23} which shows, from the geometrical point of
view, that the phase space of the electromagnetic field has NC
structure. In other words, unlike the previous approach that is
based on the comparison of the motion equations, the NC structure
of the phase space can be shown directly. Furthermore, for a
classical charged particle in an electromagnetic field, one can
display~\cite{23}, under the special condition and by a new
definition of the Poisson bracket, that the phase space variables
no longer commute in the sense of the new bracket. Indeed, it has
been demonstrated that the new bracket describes the motion
equations with respect to the new Hamiltonian~\cite{23}. This
approach is based on the Dirac theory~\cite{12} which is the
extension of the Hamiltonian formalism for a dynamical system
including constraints. Of course, it has been indicated that, in
the presence of an external background magnetic field, the NC
coordinates can naturally be introduced~\cite{11}, that is, the
(quantum) commutator of the phase space coordinates are no longer
the Poisson commutator.

We show that the Dirac theory can also be applied for the GEM
field provided that the particle velocity, say ${\bf v}$, is
sufficiently slow such that one can neglect the kinetic term in
the Lagrangian\rlap.\footnote{That is, the ratio $v^2/c^2$ is
ignored against the unity.}\
 It is in this
limit that the gravitational system can be regarded as a
constrained system, and therefore, the Dirac bracket reveals the
NC structure of the phase space.

This work is organized as follows. In the next section, we
describe the main ideas of the NC geometry needed for the work,
and give a brief formulation of the classical mechanics in this
frame. In Section~$3$, the FP and its application for the GM field
is introduced. The main purpose of this work, i.e. the equivalency
between the GM field and the NC effects, is performed in
Section~$4$. In Section~$5$, we provide a classical formulation of
the gravitoelectromagnetism based on the Dirac theory. The
conclusions are presented in the last section.

\section{Non--commutative Classical Mechanics}
\indent

The NC geometry has played an increasingly important role, more
notably, in the attempts to understand the space--time structure
at very small distances. The NC structure idea at small length
scale first introduced by Snyder~\cite{13}. He applied the NC
structure concept for discrete space--time coordinates instead of
the continuum ones. From the quantum theory point of view, the
coordinates operators of a NC space--time do~not satisfy the usual
commutation relations, i.e. the usual
$[\hat{x}^\mu,\hat{x}^\nu]=0$, and instead obey non--trivial ones.
The commutation relations are defined with respect to the notion
of $*$ (or NC) product\footnote{This product is usually called the
Weyl--Moyal product.}\
 such that for any two different
operators, e.g. ${\cal A}$ and ${\cal B}$, one has, in general,
${\cal A}*{\cal B}\neq{\cal B}*{\cal A}$.

The NC concept is~not limited only to the space--time operators,
and can be extended to the phase space classical variables as
well. In this respect, the versions of classical mechanics, e.g.
Ref.~\cite{9}, and ordinary quantum mechanics, e.g.
Ref.~\cite{10}, have been studied in terms of the NC geometry.

It is well--known that the passage from the commutative to NC
frame is simply achieved by replacing the ordinary product by the
$*$~product, as has been performed in the literature, see, e.g.,
Refs.~\cite{11}--\cite{MalkFarhoudi10}. Our approach to the NC
formalism is also based on this passage. That is, for considering
a physical theory in the NC frame, one should take its classical
version defined on a commutating frame and then, replaces the
usual product by the NC product. In the classical physics, the
non--commutativity can be described by this product, in where the
product law is defined between two arbitrary (infinitely
differentiable) functions, e.g. $f$ and $g$, of the phase space
variables (that is, $\zeta^a=(x^i,p^j)$ for $i, j=1, \cdots, n$)
as~\cite{11}
\begin{equation}\label{B1}
(f*g)(\zeta) =\exp
\Bigl[\frac{1}{2}\alpha^{ab}\partial_a^{(1)}\partial_b^{(2)}
\Bigr]f(\zeta_1)g(\zeta_2){\Bigr |}_{\zeta_1=\zeta_2=\zeta},
\end{equation}
where $a, b=1, 2, \cdots, 2n$ and $2n$ is the dimension of the
phase space. The real matrix $\alpha_{ab}$ is the generalized
symplectic structure and can be written as
\begin{equation}\label{B2}
\alpha_{ab}=\left(%
\begin{array}{cc}
\theta_{ij} & \delta_{ij}+\sigma_{ij} \\
-\delta_{ij}-\sigma_{ij} &  \beta_{ij}  \\
\end{array}%
\right).
\end{equation}
The constant antisymmetric components $\theta_{ij}$ and
$\beta_{ij}$ are called the NC parameters of the space and
momentum sectors, respectively, which can be written in terms of
the Levi--Civita antisymmetric tensor, namely
\begin{equation}\label{levi}
{\theta}_{ij}={\varepsilon_{ij}}^k{\theta}_k\qquad {\rm and}
\qquad {\beta}_{ij}={\varepsilon_{ij}}^k{\beta}_k.
\end{equation}
The real parameters $\theta_k$ and ${\beta}_k$ are usually assumed
to be very small, and hence, we consider them up to the first
order. The third parameter $\sigma_{ij}$ can be written as
combination (product) of the other two parameters and hence, can
be ignored up to the first order. Obviously, the defined
product~(\ref{B1}) is associative, but not~commutative, hence the
modified Poisson bracket can be written as
\begin{equation}\label{B3}
\{f,g\}_*=f*g-g*f.
\end{equation}
Thus, it is easy to show that
\begin{equation}\label{B31}
\{x_i,x_j\}_*=\theta_{ij},\hspace{1cm}\{x_i,p_j\}_*=\delta_{ij}+\sigma_{ij}\qquad
{\rm and} \qquad \{p_i,p_j\}_*=\beta_{ij}.
\end{equation}

A simple way to study a physical theory within the NC geometry is
the replacement of the Moyal product with the ordinary
multiplication, when one considers the following non--canonical
transformation on the classical phase space\rlap.\footnote{The
variables of the classical phase space obey the usual Poisson
brackets, namely $\{x_i,x_j\}=0=\{p_i,p_j\}$ and
$\{x_i,p_j\}=\delta_{ij}$.}\
 This transformation, which is often referred to as the
Seiberg--Witten map~\cite{9,11,CSjT01}, is
\begin{equation}\label{A7}
x'_i=x_i-\frac{1}{2}\theta_{ij}p^j\qquad {\rm and} \qquad
p'_i=p_i+\frac{1}{2}\beta_{ij}x^j.
\end{equation}
By which, the usual Poisson brackets of the primed variables give
\begin{equation}\label{A9}
\{x'_i,x'_j\}=\theta_{ij},\hspace{.5cm}\{x'_i,p'_j\}=\delta_{ij}+\sigma_{ij}\qquad
{\rm and} \qquad \{p'_i,p'_j\}=\beta_{ij},
\end{equation}
with $\sigma_{ij}=-\theta_{k(i}\beta_{j)l}\delta^{kl}/4$. The
commutation relations (\ref{A9})  are the same as (\ref{B31}), and
consequently, for introducing non--commutativity, it is more
convenient to work with the Poisson brackets (\ref{A9}) than the
modified brackets (\ref{B31}). It is important to note that, the
relations represented by equations (\ref{B31}) are defined in the
spirit of the Moyal product given above, however, in the relations
defined by (\ref{A7}) and used in (\ref{A9}), the variables $(x_i,
p_j)$ obey the usual Poisson bracket. Hence, the two sets of the
deformed and ordinary Poisson brackets must be considered as
distinct.

We assume that one has a symplectic structure consistent with the
commutation rules (\ref{A9}) and then, obtain the corresponding
equations of motion, where it can be understood as the motion
equation of a particle in the NC space. Also, it is assumed that
the functional form of the Hamiltonian in the commutative and NC
cases are in the same form, i.e.
\begin{equation}\label{B4}
H(x',p')=\frac{p'^2}{2m}+V(x'),
\end{equation}
where the coordinates $x'_i$ and the momenta $p'_i$ ($i=1,2,3$)
yield brackets (\ref{A9}). The Hamiltonian equations are
\begin{equation}\label{B55}
\dot{x'}_i=\{x'_i,H\}\qquad\quad {\rm and}\qquad\quad
\dot{p'}_i=\{p'_i,H\},
\end{equation}
which easily govern the NC dynamics, up to the first order, as
\begin{eqnarray}\label{B5}
\dot{x'}_i&=&\frac{p'_i}{m}+\theta_{ij}\frac{\partial
V}{\partial x'_j}+\frac{\sigma_{ij}p'^j}{m}\simeq
\frac{p'_i}{m}+\theta_{ij}\frac{\partial V}{\partial x'_j},\nonumber\\
\dot{p'}_i&=&-\frac{\partial V}{\partial
x'^i}+\frac{\beta_{ij}p'^j}{m}-\sigma_{ij}\frac{\partial
V}{\partial x'_j}\simeq -\frac{\partial V}{\partial
x'^i}+\frac{\beta_{ij}p'^j}{m}.
\end{eqnarray}
By eliminating the momentum variables, one gets
\begin{equation}\label{A3}
 m{\ddot{x'}}_i\simeq-\frac{\partial V}{\partial
 x'^i}+\Bigl[m\theta_{ij}\Bigl(\frac{\partial^2
 V}{\partial{x'}^k\partial{x'}_j}\Bigr)+{\beta}_{ik}\Bigr]{\dot{x'}}^k.
\end{equation}
The first term in the right hand side~(\ref{A3}) is the usual
conservative force. The second and third terms, which depend on
the velocity, are arisen from the non--commutativity and can be
interpreted as additional forces. It means that, the NC effects
can be equivalent with imposing the additional forces on the usual
(commutative) space.

In the case of a free particle, as the simplest example, equation
(\ref{A3}) reads
\begin{center}
$m{\ddot{x'}}_i\simeq {\beta}_{ik}{\dot{x'}}^k
={\varepsilon}_{ikj}
{\dot{x'}}^k\beta^j=(\textbf{v}'\times\mbox{\boldmath
$\beta$})_i$,
\end{center}
where we take $\mbox{\boldmath $\beta$}=(\beta^j)$. This equation
is similar to the motion equation of a classical moving charged
particle $q$ in the presence of a magnetic field ${\bf B}_{\rm
m}$, i.e. $m{\ddot{x}}_i=q(\textbf{v}\times{\bf B}_{\rm m})_i$ in
the SI units. Thus, the effect of the NC parameters $\beta^j$ in
the NC space is similar to the presence of a magnetic field in the
usual space. The comparison between these two equations, gives the
magnitude of the NC parameter corresponding to the magnetic field,
namely
\begin{equation}\label{A4}
\textbf{B}_{\rm m}=\mbox{\boldmath $\beta$}/q.
\end{equation}
Indeed, equation (\ref{A4}) represents the equivalency of the NC
effects and the magnetic force field.

The above example motivates the following issue. As the GM field
is the analogous of the magnetic field in the
gravitoelectromagnetism, there would also be the corresponding
equivalency of the GM field and the NC parameters. In the explicit
words, one expects that the NC effects can be interpreted as the
impose of the GM field in the usual space. Equivalently, it is
interesting to find out what is the relevance of the
gravitomagnetism from the NC geometry effects. In the next two
sections, we probe this modification, though first, in
Section~$3$, we investigate the Lorentz--like behavior of the GM
field.

\section{Gravitomagnetism and Feynman Proof}
\indent

In the absence of the electromagnetic interactions, the geodesic
equation for particles position is obviously valid. For instance,
in the particular case of a rotating sphere, when the space--time
metric is given by the Kerr metric in the Boyer--Lindquist
coordinates~\cite{15}, the gravitational Lorentz force can easily
be obtained by using the geodesic equation~\cite{16}
$\ddot{x}^\lambda+\Gamma^\lambda{}_{\mu\nu}\dot{x}^\mu\dot{x}^\nu=0$.
In other words, the existence of a Lorentz--like force law is also
possible for the gravitational interactions under the special
conditions. Indeed, Feynman gave the proof of the Maxwell
equations independent of the electromagnetic theory, just by
assuming the Newtonian law of motion and the commutation relation
between the position and the velocity for a single
non--relativistic particle. The details of this approach have been
presented in the Dyson work~\cite{17}. It is important to
emphasize that the proof reproduces only the homogeneous Maxwell
equations (i.e. the two free source equations), which are
compatible with the Galilean relativity~\cite{vaidya91}.

Let us make a concise review on the FP without mathematical
rigorous proof of the theorem, though for a complete
demonstration, see, e.g., Refs.~\cite{101,18,20}. Incidentally, we
consider the FP of the Maxwell equations in the classical form and
follow the Dyson approach~\cite{17}.

The proof takes the classical and quantum concepts for a single
non--relativistic particle, and leads to the Lorentz--like force
law and the homogeneous Maxwell equations. The result seems
somehow strange, for by starting with a classical equation (the
Newtonian law), one will end up with the relativistic one. The
basic assumptions of the proof consist of
\begin{itemize}
\item{The Newtonian second law},
\end{itemize}
\begin{itemize}
\item{The Galilean relativity},
\end{itemize}
\begin{itemize}
\item{The commutation relations between the position and the velocity of
a particle}.
\end{itemize}

Consider a particle whose the position $x_i$ and the velocity
$\dot{x_i}$ satisfy the Newtonian second law
\begin{equation}
m\ddot{x_i}=F_i(x,\dot{x},t),
\end{equation}
and also obey the commutation relations\footnote{In the case of
classical mechanics, and when the equations of motion are
canonical, the commutation relations can be considered as the
usual Poisson brackets~\cite{101,18}.}
\begin{equation}
[x_i,x_j]=0\qquad\quad {\rm and}\qquad\quad
m[x_i,\dot{x}_j]=\delta_{ij}.
\end{equation}
Hence, it can be shown that there exist an $\textbf{E}(x,t)$ field
and a $\textbf{B}(x,t)$ field such that
\begin{equation}\label{F3}
F_i=E_i+\varepsilon_{ijk}\dot{x}^jB^k,
\end{equation}
with
\begin{equation}\label{F19}
\mbox{\boldmath $\nabla$}\cdot\textbf{B}=0\qquad\quad {\rm
and}\qquad\quad \mbox{\boldmath
$\nabla$}\times\textbf{E}+\frac{\partial \textbf{B}}{\partial
t}=0,
\end{equation}
in the SI units. As it is obvious, this theorem not~only includes
the electromagnetic phenomena, but also discusses more general
cases, even it can be extended to the case of non--Abelian gauge
fields, both in the Newtonian~\cite{Lee90} as well as in the
relativistic dynamics~\cite{101} and the dynamical equations of
spinning particles~\cite{20}. Also, it can be shown that the FP is
applicable to the GEM phenomena in the case of stationary
fields~\cite{21}\rlap.\footnote{In the non--stationary case, the
electric--like and the magnetic--like fields in equation
(\ref{F3}) do~not satisfy the two Maxwell--like
equations~\cite{211}.}

\section{Gravitomagnetic Field and Non--commutative Effects}
\indent

In the above discussion, we consider a moving mass particle in the
gravitational field when the stationary weak field and the slow
motion approximations are satisfied. Thus, equation (\ref{F3}) can
be viewed as the motion equation of the particle in the
gravitational field, and hence, the first and second terms in the
right hand side of (\ref{F3}) exhibit the gravitoelectric and the
GM forces, respectively. That is, equation (\ref{F3}) can be
re--written in the form
\begin{equation}\label{B13111}
m\ddot{x_i}=-\frac{\partial V}{\partial x
^{i}}+m\varepsilon_{ijk}\dot{x}^{j}B_{\rm gm}^{k},
\end{equation}
where $E_{i}=-\partial V/\partial x ^{i}$ (with $V$ as the
Newtonian weak potential), and $B_{\rm gm}^{i}=B^{i}/m$ is the
corresponding GM field.

On the other hand, in the case of stationary fields, potentials
are linear in terms of the coordinates, and if one substitutes
such a potential in the NC motion equation (\ref{A3}), the second
term will be dropped, and one attains
\begin{equation}\label{B132}
m\ddot{x_i}=-\frac{\partial V}{\partial x
^{i}}+\varepsilon_{ijk}\beta^{k}\dot{x}^{j}.
\end{equation}
Though we have dropped the primes, but we should remind that this
equation describes the equation of motion of a particle in the NC
space. By comparing the second terms on the right hand sides of
equations (\ref{B13111}) and (\ref{B132}), one is led to the
following statement:

\vspace{3mm}
\noindent\emph{When the geometry of the phase space is NC, the
force term, that comes from the non--commutativity of the momentum
sector, is equivalent to the exertion of the GM force in the usual
space.}

\vspace{3mm}
Also, if one formally equates these terms, one will get
\begin{equation}\label{B133}
\textbf{B}_{\rm gm}=\mbox{\boldmath $\beta$}/m,
\end{equation}
which represents the gravitational counterpart of equation
(\ref{A4}). Therefore, the NC effects of the momentum sector is
analogous to the presence of the GM field. Another and a more
accurate statement of the above result can be expressed as
follows:

\vspace{3mm}
\noindent\emph{The effect of NC parameter of the momentum sector
for a static mass distribution in a NC space is equivalent to when
the same mass distribution is in uniform motion in the usual
space.}

\vspace{3mm}
The following thought experiment, as a simple example, can be
instructive and, lends more justification for the subject.

Suppose, in the usual space, a constant distribution of mass
surface density $\sigma$, say, in the $xy$--plane, is moving in
the $x$--direction with constant speed $u$. With the calculation
of the corresponding electromagnetic problem~\cite{2111}, or the
direct calculation~\cite{21111}, one can show that the
gravitational Lorentzian force on a mass particle $m$ that moves
in the $x$--direction is
\begin{center}
$\textbf{F}=\textbf{F}_{\rm ge}+\textbf{F}_{\rm gm}=-2\pi Gm
\sigma \hat{\textbf{z}}+4\pi Gm\sigma u v_x\hspace{1mm}
\hat{\textbf{z}}/c^2$,
\end {center}
where $v_x$ is the $x$--component of the particle velocity and
$\textbf{F}_{\rm ge}$ and $\textbf{F}_{\rm gm}$ are the
gravitoelectric and the GM forces, respectively. In accord with
equation (\ref{B13111}), the GM field can easily be found to be
\begin{center}
$\textbf{B}_{\rm gm}=4\pi G\sigma u\hspace{1mm}
\hat{\textbf{y}}/c^2$.
\end {center}
By equation (\ref{B133}), the NC parameter $\beta$ can be
described in terms of the statical and the kinematical quantities
of the subject, namely
\begin{center}
$\beta=4\pi Gm\sigma u/c^2$.
\end {center}

Note that, in the NC space, the mass distribution (in the
$xy$--plane) is in the rest, but still the same force acts on the
particle. This fact is concealed in the dependency of the NC
parameter on the velocity of the mass distribution.

\section{Gravitomagnetism and Dirac Bracket}
\indent

Canonical variables of the phase space obey the usual Poisson
brackets. However, we have shown that when one considers
transformation (\ref{A7}) on the usual phase space, then the
Poisson brackets of the new (prime) variables obey the new
brackets (\ref{A9}). In geometrical language, the Poisson brackets
are mapped into the modified brackets through this transformation.
Hence, one obtains the motion equation in the NC space, and can
match it with those of the usual space in the presence of a GM
field. This leads to the equivalency of the NC effects and the GM
force field. Now, in this section, we illustrate that, without
refereing to the NC space (and hence, without appealing to the
coordinates transformation (\ref{A7})), one can define the new
brackets on the phase space of the GEM field, and shows directly
that the geometry of the phase space in the scene of these new
brackets is actually NC. Of course, defining a new set of brackets
can be considered as a typical transformation of the Poisson
brackets. From the geometrical aspect, it means that the Poisson
structure defined on the manifold\footnote{Here, the underlying
manifold is the phase space.}\
 is actually replaced by a new algebraic structure known as the Dirac
structure (brackets).

Indeed, for constrained dynamical systems, Dirac introduced his
brackets instead of the Poisson brackets, and as we will show, the
GEM system can be considered as a constrained system. Hence, one
just needs to apply the Dirac canonical formalism to proceed the
above issue.

Dirac formalism is actually an extension of the Hamiltonian
formulation for mechanical systems which are subject to
constraints. That is, although the motion equations of systems
without constraints are expressible in terms of the Poisson
brackets, but for systems involving with constraints, the
Hamiltonian equations of motion can be expressed in terms of the
Dirac brackets. And, a constrained system is a system whose
position and momentum variables obey certain identities
(constraints) and therefore, the canonical variables are~not
independent. Of course, such a dependency comes from the form of
the Lagrangian function, which is often referred to as the
nonstandard or the degenerate Lagrangian. Indeed, Dirac extended a
standard technique and used a new Hamiltonian form for the
extension of such nonstandard Lagrangians. For a more
justification, let us look at the subject from the following point
of view.

In the usual formulation of classical mechanics, the passage from
the Lagrangian variables of generalized positions $q_i$ and
velocities $\dot{q}_i$, to the Hamiltonian variables (i.e.
generalized positions and momenta $p_i$) is possible when, and
only when, the velocities can be expressed in terms of the
positions and momenta. This is done by the Legender
transformation, i.e. $(q_i, \dot{q}_i)\rightarrow (q_i, p_i)$,
where $ p_i=\partial L(q,\dot{q})/\partial \dot{q}^i$ and
$L(q,\dot{q})$ is the Lagrangian, see, e.g.,
Ref.~\cite{goldstein}. It is expected that the definition of
momenta should~not lead to any identities among the positions and
momenta alone, for the canonical variables must be independent.
But, there are some situations for which the definition of
generalized momenta leads to such identities or, in an another
word, constraints. An explicit example of such situation is when
the Lagrangian is linear in generalized velocities. Thus, the
transition from the Lagrangian to the Hamiltonian formalism at the
classical level is non--trivial. This transition and also the
canonical quantization of such constrained systems were worked out
by Dirac~\cite{12}. Both the Poisson bracket and commutator are
representations of a Lie algebra product, hence the corresponding
principle, in replacing the classical Poisson bracket by the
commutator of the quantum mechanical operators, can work. Dirac
introduced a procedure for the Hamiltonian formalism of a
constrained system, and considered the new Hamiltonian form for
developing the quantum mechanics of such systems. Of course, the
classical version of the Dirac theory has been well studied in the
literature, see, e.g., Ref.~\cite{22}. In below, we give a brief
review on account of the Dirac theory in the classical version
while avoiding the mathematical rigorous\rlap.\footnote{For a more
complete discussion of this issue see, e.g.,
Refs.~\cite{goldstein,221,222}.}

Let $x_1,\ldots ,x_n$ and $y_1,\ldots ,y_n$ be a coordinate system
on $2n$ dimensional Euclidean space $\textsf{R}^{2n}$ and let
$\textsf{U}$ be an open subset of $\textsf{R}^{2n}$. Then, the
Poisson bracket of any two infinitely differentiable functions on
$\textsf{U}$, say $f$ and $g$, is defined and denoted by
\begin{equation}
\{f,g\}=\sum_{i=1}^n\Big(\frac{\partial f}{\partial
x_i}\frac{\partial g} {\partial y_i}-\frac{\partial f}{\partial
y_i}\frac{\partial g}{\partial x_i}\Big).
\end{equation}
The symplectic form $\omega$, corresponding to the Poisson
bracket, is called the classical symplectic structure and is
represented by the $2n\times 2n$ matrix
\begin{equation}
\omega=\left(
         \begin{array}{cc}
          \hspace{0.34 cm}\textbf{0} \hspace{1 cm}\,\,          \textbf{1}_{n\times n} \\
           \!\!\!-\textbf{1}_{n\times n} \hspace{1 cm}        \textbf{0} \\
         \end{array}
       \right)
\end{equation}
such that, the Poisson bracket can be written in the form
\begin{equation}
\{f,g\}=\pi^{ij}\partial_if \partial_jg,
\end{equation}
where $\pi^{ki}\omega_{jk}=\delta^i_j$.

For mechanical systems, $\textsf{R}^{2n}$ can be considered as
$2n$ dimensional phase space with the canonical variables
$(q_1,\ldots ,q_n$ and $p_1,\ldots ,p_n)$. On the other hand, in
many cases, constraints can be written in terms of the phase space
functions. In this manner, let $\phi_i(q, p)=0$  ($i=1,\ldots ,m$)
denotes all constraints for the Hamiltonian system. These
constraints can be divided into two classes by analyzing the
$m\times m$ antisymmetric matrix of their mutual Poisson brackets,
say $M_{ij}=\{\phi_i, \phi_j\}$. Since $M_{ij}$ is antisymmetric,
its rank $r$ must be even\rlap,\footnote{The rank of a matrix is
the number of its independent row (column).}\
 and after
redefining the constraints by taking their linear combinations
(known as the Dirac separating constraints algorithm), the top
left $r\times r$ sub--matrix of $M_{ij}$, which we denote by
$C_{ij}$, is regular (i.e. nonsingular). The constraint functions
$\phi_{r+1},\ldots ,\phi_m$ are called the first--class and
$\phi_1,\ldots ,\phi_r$ are called the second--class. Regarding
the definition of Dirac bracket, the second--class constraints are
only considered, and for them, we can introduce the Dirac bracket
of any two phase space functions $f$ and $g$ as
\begin{equation}
\{f,g\}_D=\{f,g\}- \{f,\phi_i\}C^{ij}\{\phi_j,g\},
\end{equation}
where the matrix $C^{ij}$ is the inverse of $C_{ij}$. Note that,
the functions $\phi_i(q,p)$ are a certain subset (second--class)
of all those functions whose vanishing give the constraints. Dirac
argued that one can generalize the canonical Hamiltonian $H$ to
the new Hamiltonian $H_T$ (the total Hamiltonian) such that, the
dynamics is confined by the constrains and is governed by the
total Hamiltonian. Namely,
\begin{equation}
\dot{q}_i=\{q_i, H_T\}_D\qquad\quad{\rm and}\qquad\quad
\dot{p}_i=\{p_i, H_T\}_D
\end{equation}
with
\begin{equation}\label{BB0}
H_T=H+\sum_{k=1}^m u^k\phi_k,
\end{equation}
where $u^k$'s are unknown coefficients and are~not necessarily
functions of the coordinates and momenta. Recall that, when the
constraints are exerted, one gets $H_T=H$. Thus, a constrained
Hamiltonian system is defined by a given Lagrangian together with
the Dirac bracket corresponding to the set of constraints. The
physical phase space of the system is a constraint manifold, which
is a sub--manifold of the unconstraint (naive) phase space
$\textsf{R}^{2n}$ spanned by the $p_i$'s and $q_i$'s. An another
(short) statement of the above discussion can be asserted as
follows.

The flat symplectic (Poisson) structure on the naive phase space
induces a non--trivial symplectic structure on the physical phase
space. This non--trivial symplectic structure is given by the
Dirac bracket and it is just what can show the non--commutativity
in the physical phase space.

Let us state the result of an example\footnote{The full
illustration of this example can be found in
Ref.~\cite{sgavery}.}\
 in the electromagnetism for the
motion of a charged particle $q$ with mass $m$ in a constant
magnetic field $B_0$. The motion is subjected to the following
condition which we refer to it as the magnetic condition:

\vspace{3mm}
\noindent\textsl{The ratio $qB_0/2mc$ is sufficiently large such
that the kinetic term in the corresponding Lagrangian can be
neglected.}

\vspace{3mm}
\noindent In this example and under the magnetic condition, the
Dirac brackets show the non--commutativity in the phase space of
the electromagnetic field. Evidently, this condition induces a
strong magnetic field\rlap,\footnote{Though, a considerable amount
of charge $q$ or a very small mass $m$ can also make this large
ratio, but we do~not consider these cases.}\
 while in the corresponding
gravitoelectromagnetism, the GM field is a weak field. However,
one can still employ the Dirac theory for the GEM field without
the magnetic condition, but when the particle, that moves in a
constant GM field, is subjected to the GM condition as:

\vspace{3mm}
\noindent\textsl{The velocity of particle, ${\bf v}$, is so small
that the ratio $v^2/c^2$ can  be neglected against the unity.}

\vspace{3mm}
Indeed, this GM condition is a typical statement of the slow
motion condition, i.e. the smallness of the particle velocity in
contrast to the light velocity. And, it is well--known that the
slow motion is one of the conditions that general relativity
reveals the electromagnetism\rlap.\footnote{Obviously, the full
conditions are the weak field and the slow motion
approximations.}\
 It looks that the magnetic condition and the GM condition are another
dissimilarity between the GM field and the magnetic field, but, in
Ref.~\cite{23}, it has been shown when the magnetic field is weak,
the non--trivial Dirac bracket can also be resulted. This apparent
limitation of weak magnetic fields is even a desirable feature of
the approach of Ref.~\cite{23}, where it has been proved that by a
given set of Dirac constraints, a magnetic field induces the
corresponding Dirac brackets. In fact, by extending the usual
understanding of the classical phase space and regarding it as a
generalized complex manifold, a unified picture is obtained in
which the magnetic fields and the Dirac constraints turn out to be
equivalent objects. In below, analogous to the electromagnetic
example, we illustrate that the latter statement is also true for
the GM field.

The Lagrangian for a particle of mass $m$ and velocity ${\bf v}$
moving in a GEM field is given by~\cite{24}
\begin{equation}\label{BB1}
L=-mc^2(1-v^2/c^2)^{1/2}+m\frac{1+v^2/c^2}{\sqrt{1-v^2/c^2}}U-\frac{2m/c}{\sqrt{1-v^2/c^2}}\mathbf{A}\cdot\mathbf{v},
\end{equation}
where $U$ and $\mathbf{A}$ are the gravitoelectric and the GM
potentials, respectively. By ignoring $v^2/c^2$ against the unity,
multiplying by $1/m$ and discarding a constant, the Lagrangian can
be written in the form
\begin{equation}\label{BB2}
L\approx U-\frac{2}{c}\mathbf{A}\cdot\mathbf{v}.
\end{equation}
Lagrangian (\ref{BB2}) is linear in terms of velocity and
therefore, the equations of the generalized momenta can be
considered as the Dirac constraints. It is instructive to
re--scale the potential as $U\rightarrow \epsilon\hspace{.3mm} U$
with $\epsilon=B_{\rm gm}/c$, where we assume that the particle is
confined to the $xy$--plane in a constant GM field ${\bf B}_{\rm
gm}=B_{\rm gm}\hat{\mathbf{z}}$. Therefore, Lagrangian (\ref{BB2})
reads
\begin{equation}\label{BB3}
L\approx \epsilon \hspace{.3mm}U-\epsilon(x\dot{y}-y\dot{x}),
\end{equation}
where the GM vector potential has been considered to be
${\mathbf{A}}=-{\mathbf{r}}\times {\mathbf{B}}_{\rm gm}/2$. The
conjugate momenta are obtained from Lagrangian (\ref{BB3}) as
\begin{equation}\label{BB4}
p_x=\epsilon \hspace{.4mm}y\qquad\quad{\rm and}\qquad\quad
p_y=-\epsilon \hspace{.4mm}x,
\end{equation}
and thus, the corresponding Hamiltonian takes the simple form
\begin{equation}\label{BB5}
H=-\epsilon \hspace{.3mm}U.
\end{equation}
Equations (\ref{BB4}) give two primary constraints, namely
\begin{equation}\label{BB6}
\phi_1=\epsilon\hspace{.3mm} y-p_x\qquad\quad{\rm and}\qquad\quad
\phi_2=\epsilon \hspace{.3mm}x+p_y.
\end{equation}
Therefore, according to definition (\ref{BB0}), the total
Hamiltonian is
\begin{equation}\label{B13}
H_T=-\epsilon\hspace{.2mm} U+u_1(\epsilon
\hspace{.3mm}y-p_x)+u_2(\epsilon \hspace{.3mm}x+p_y),
\end{equation}
and hence, we have
\begin{equation}\label{B131}
\{\phi_1, H_T\}=\epsilon(2u_2-\frac{\partial U}{\partial
x})\qquad\quad{\rm and}\qquad\quad \{\phi_2,
H_T\}=-\epsilon(2u_1-\frac{\partial U}{\partial y}).
\end{equation}
Note that, for consistency~\cite{12}, when the constraints are
imposed, one must set $\{\phi_i,H_T\}=0$ ($i=1, 2$).

The latter equations are regarded as the relations that determine
the coefficients $u_1$ and $u_2$ in (\ref{B13})\rlap.\footnote{For
this reason, these relations are usually called the consistency
conditions.}\
 Also, it can easily be found that
\begin{equation}
\{\phi_1,\phi_2\}=-\{\phi_2,\phi_1\}=2\epsilon,
\end{equation}
as expected not~to be zero, for our primary constraints are the
second--class ones. Thus, one can define the Dirac bracket in
terms of these constraints. Let us constitute the matrix
$C_{ij}=\{\phi_i,\phi_j\}$, that is
\begin{equation}
(C_{ij})=-2\epsilon\left(
  \begin{array}{cc}
    0 & -1 \\
    1 & 0 \\
  \end{array}
\right),
\end{equation}
whose inverse is
\begin{equation}
(C^{ij})=-\frac{1}{2\epsilon}\left(
  \begin{array}{cc}
    0 & 1 \\
    -1 & 0 \\
  \end{array}
\right).
\end{equation}
The inverse matrix elements can be written as\footnote{The
notation $\varepsilon^{ij}$ is the Levi--Civita symbol in two
dimensions.}\
 $C^{ij}=-\varepsilon^{ij}/2\epsilon$,
hence, the Dirac bracket of any two quantities $f$ and $g$ is
\begin{equation}\label{B130}
\{f, g\}_D=\{f,
g\}+\frac{1}{2\epsilon}\varepsilon^{ij}\{f,\phi_i\}\{\phi_j, g\}.
\end{equation}

Now, the non--trivial Dirac bracket of the phase space coordinates
can be resulted. However, before this, it will be an instructive
and useful example, if one writes the motion equation in terms of
the Dirac bracket. For this purpose, equation (\ref{B130}) for $x$
variable gives
\begin{eqnarray*}
\dot{x}=\{x, H_T\}_D=u_1\{x,
\phi_1\}+\frac{1}{2\epsilon}\{x,\phi_1\}\{\phi_2, H_T\} =u_1\{x,
\phi_1\}-\frac{1}{2\epsilon}\{x,\phi_1\}\epsilon(2u_1-\frac{\partial
U}{\partial y}),
\end{eqnarray*}
where by using (\ref{BB6}) and hence, substituting for
$\{x,\phi_1\}=-1$, leads to
\begin{eqnarray*}
\dot{x} =-\frac{1}{2}\frac{\partial U}{\partial y}.
\end{eqnarray*}
For $y$ variable, we get
\begin{eqnarray*}\label{B134}
\dot{y}=\{y, H_T\}_D=u_2\{y,
\phi_2\}-\frac{1}{2\epsilon}\{y,\phi_2\}\epsilon(2u_2-\frac{\partial
U}{\partial x}),
\end{eqnarray*}
that as $\{y,\phi_2\}=1$, gives
\begin{eqnarray*}\label{B135}
\dot{y}=\frac{1}{2}\frac{\partial U}{\partial x}.
\end{eqnarray*}
With a similar calculation, the equation motions of momenta are
\begin{eqnarray*}\label{B136}
\dot{p_x}=\{p_x, H_T\}_D=\frac{1}{2}\epsilon\frac{\partial
U}{\partial x}\qquad\qquad{\rm and}\qquad\qquad \dot{p_y}=\{p_y,
H_T\}_D=\frac{1}{2}\epsilon\frac{\partial U}{\partial y}.
\end{eqnarray*}
However, the usual Hamilton equations give
\begin{eqnarray*}\label{B138}
\dot{x}=0=\dot{y},
\end{eqnarray*}
which are~not the correct results, and also
\begin{eqnarray*}\label{B1388}
\dot{p_x}=\epsilon\frac{\partial U}{\partial x}\qquad\qquad{\rm
and}\qquad\qquad  \dot{p_y}=\epsilon\frac{\partial U}{\partial y},
\end{eqnarray*}
which contradicts the outcomes of the constrain equations
(\ref{BB4}). These results confirm the necessity of the new
Hamiltonian $H_T$.

Let us return to the main purpose of the work. It is easily to
show that the Dirac brackets of the phase space coordinates are
\begin{equation}\label{B139}
\{x, y\}_D=\{x,
y\}+\frac{1}{2\epsilon}\varepsilon^{ij}\{x,\phi_i\}\{\phi_j,
y\}=\frac{1}{2\epsilon}\{x,\phi_1\}\{\phi_2,
y\}=\frac{1}{2\epsilon}.
\end{equation}
Similarly, we get
\begin{equation}\label{B140}
\{x, p_x\}_D=\frac{1}{2}=\{y, p_y\}_D,
\end{equation}
and
\begin{equation}\label{B141}
\{p_x, p_y\}_D=\frac{\epsilon}{2}.
\end{equation}
Obviously, the Dirac brackets between the phase space coordinates
are nonzero, particularly for the momentum sector. It can be
resulted that, the geometry of the phase space is NC and the
associated parameter of the momentum sector is proportional to the
magnitude of the GM field. In other words, the non--commutativity
in the momentum sector is equivalent to the presence of the GM
force whose weakness implies that the NC parameter of the momentum
(space) sector is small (large).

\section{Conclusions}
\indent

Gravitomagnetism is an old subject in general relativity around
which many well written papers exist. However, the effects of
accompanying non--commutativity lends some weight to its
significance. On the other hand, the generalized uncertainty
principle which is somehow what we have considered in this study
has gained a considerable amount of momentum in recent years and
is thus carries a noticeable importance. However, its use in the
framework of gravitomagnetism has been somewhat left on the side.
It is thus interesting to study within this framework. Hence, in
this work, we have investigated one of the other similarities
between the GM and the magnetic fields. The investigation is
relevant to the equivalence between the GM field and the NC space
effects. For this purpose, we have proceeded through two
approaches. The first approach is based on the comparison of the
motion equations of a particle in a NC space and the usual space
in the presence of a GM field. In the usual space, according to
the FP, the motion equation includes a GM term that comes from the
motion of the potential source. Indeed, the presence of such term
is resulted from the existence of the Lorentz--like force for the
GEM field. The existence of such force is provided by the slow
motion and the weak field approximations when the GEM fields are
in the stationary case. The comparison of the two motion equations
shows that, the GM field (like its magnetic analogous) can be
equivalent to the NC parameter of the momentum sector. That is,
the NC effect can be interpreted as the presence of the GM field
in the usual space.

The second approach is based on the fact that the GEM system can
be considered as a constraint system in the Dirac formalism. The
weakness of the field and sufficiently low velocity conditions
permit the GEM system to be described in terms of the degenerate
Lagrangian which is characterized by the constraints. Such
constraints are divided into two classes, the first-- and the
second--classes. The first--class constraints are correlated, at
least to some extent, with the gauge properties of the system,
while the second--class constraints reflect the appearance of non
dynamical degrees of freedom in the theory. The dynamics of a
constrained system is described by the Dirac bracket which is a
generalization of the Poisson bracket in the symplectic mechanics.
By the definition of Dirac bracket in terms of the second--class
constraints, the dynamics is governed by the total (extended)
Hamiltonian instead of the canonical one. With respect to the new
(algebraic) symplectic structure, the phase space variables no
longer commute and this illustrates the NC structure of the phase
space geometry. Particularly, the Dirac bracket of the momentum
sector variables is non--trivial and proportional to the magnitude
of the GM field. The main conclusion of the second approach can be
summarized in the following statement:

\vspace{3mm}
\noindent\emph{The GM field as a typical weak field is equivalent
to the Dirac constraint which induces the corresponding Dirac
bracket.}
%%%%%%%%%%%%%%%%%%%%%%%%%%%%%%%%% References %%%%%%%%%%%%%%%%%%%%%%%%%%%%%%%%%%

%
\end{document}